\documentclass[12pt]{iopart}
\usepackage{cite} 
\usepackage{graphicx}
\usepackage{amsmath}
\begin{document}

\title[Kuramoto oscillators in random networks]{Kuramoto oscillators in random networks}

\author{Agostino Funel}

\address{ENEA - P.le E. Fermi 1, 8055 Portici (NA), Italy}
\ead{agostino.funel@enea.it}
\vspace{10pt}
%\begin{indented}
%\item[]November 2024
%\end{indented}

\begin{abstract}
By means of numerical analysis conducted with the aid of the computer, the collective synchronization of coupled phase oscillators in the Kuramoto model in the connected regime of random networks of various sizes is studied. The oscillators synchronize and achieve phase coherence, and this process is not significantly affected by the level of connectivity of the network. If the probability that two oscillators are coupled is around the network connectivity threshold synchronization still occurs, although in a more attenuated way. If the size of the network is sufficiently large the oscillators have a phase transition.
\end{abstract}

%
% Uncomment for keywords
\vspace{2pc}
\noindent{\it Keywords}: Complex networks, Kuramoto model, synchronization 
%
% Uncomment for Submitted to journal title message
%\submitto{\JPC}
%
% Uncomment if a separate title page is required
%\maketitle
% 
% For two-column output uncomment the next line and choose [10pt] rather than [12pt] in the \documentclass declaration
%\ioptwocol
%

\section{Introduction}
The phenomenon of collective synchronization of an ensemble of interacting entities can be observed in different areas. For example in biology, in processes involving the regulation of the cell cycle~\cite{Banfalvi_2011,Manukyan_2011}; in physics, in the study of coupled lasers\cite{Ziping_1993,Kourtchatov_1995,Takemura_2021} and Josephson junctions\cite{Wiesenfeld_1998}; but also in human~\cite{Shahal_2020} and ecological networks~\cite{Vandermeer_2021}. One of the first important works on the subject was written by A. Winfree~\cite{Winfree_1967} in the 1960s who, studying a population of interacting oscillators, discovered that if the natural frequencies of the oscillators were very close, there existed a threshold condition for the coupling constant that, if exceeded, caused a phase transition in which some oscillators suddenly synchronized. Winfree's ideas were taken up by Kuramoto~\cite{Kuramoto_1984} who proposed a model whose analytical solution could be studied in the mean-field approximation. The equations of dynamics, for a collection of $N$ oscillators, are:

\begin{equation}
\label{eq_1_kuramoto}
\dot\theta_i = \omega_i + \frac{K}{N} \sum_{j=1}^{N} \mbox{sin}(\theta_j - \theta_i)\,.
\end{equation}

\begin{figure}[htbp]
\centering 
\includegraphics[scale=0.5]{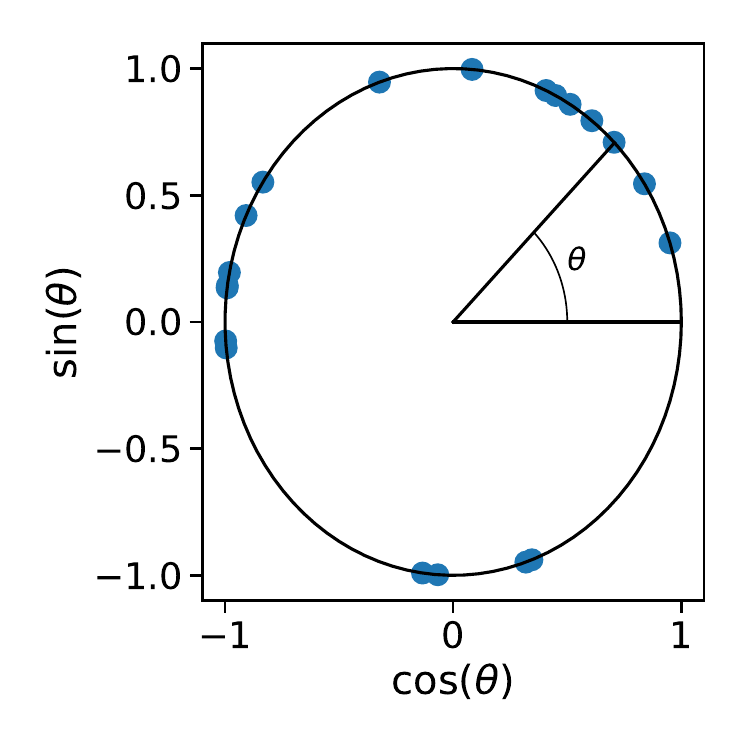}
\caption{Geometric representation of oscillator dynamics in the Kuramoto model. Motion occurs on a circle of unit radius. The azimuthal angle $\theta$ represents the phase of the oscillator.}
\label{fig1_kuramoto_phases}
\end{figure}

The temporal evolution of the phase $\theta_i$ of each oscillator depends on its natural frequency $\omega_i$, and on the phases of the other oscillators with which it interacts. If $\omega_i = \omega_j \,\, \forall i,j$ the model is called homogeneous, otherwise inhomogeneous. The magnitude of the interaction is given by the coupling constant $K \geq 0$, which is the same for all oscillators. In order to have a geometric representation of the dynamics, it is convenient to imagine that the motion occurs on a circle of unit radius so that the phase of an oscillator is represented by the azimuthal angle defined by its position, see Fig.~\ref{fig1_kuramoto_phases}. 

Collective motion can be studied by introducing the order parameter

\begin{equation}
\label{eq_2_orderpar}
R e^{i \psi} = \frac{1}{N} \sum_{j=1}^{N} e^{i \theta_j}\,.
\end{equation}

The modulus $0 \leq R \leq 1$ measures phase coherence. If the oscillators are scattered along the entire circle then $R \approx 0$  while if they are in phase coherence ($\theta_i \approx \theta_j, \forall i,j$) then $R \approx 1$. The angle $\psi$ is the average phase. If $K = 0$  the oscillators decouple and the phase of each of them evolves according to the equation $\theta_i(t) = \omega_i t + \theta_i(0)$ and synchronization, that is $\dot\theta_i = \dot\theta_j = \dot\theta$ $\forall i,j$, can only occur if $\omega_i = \omega_j \forall i,j$. Note that by summing the members of the Eq.(~\ref{eq_1_kuramoto}) we obtain $\sum_{i=1}^{N}\dot\theta_i = \sum_{i=1}^{N} \omega_i$, which implies that if all the oscillators synchronize they will have a frequency equal to the average of the natural frequencies: $\dot\theta = \bar\omega = N^{-1} \sum_{i=1}^{N} \omega_i$. In terms of the order parameter Eq.(~\ref{eq_1_kuramoto}) can be written as:

\begin{equation}
\label{eq_3_kura_op}
\dot\theta_i = \omega_i + K R \, \mbox{sin}(\psi - \theta_i) 
\end{equation}

where it is evident how the single oscillator can be thought as interacting with a mean-field with effective coupling constant equal to $K R$ and phase $\psi$.
The numerical analysis of the Kuramoto model shows the existence of a threshold value of the coupling constant below which the oscillators evolve while remaining in a state of phase incoherence; gradually increasing $K$ beyond this threshold the incoherent state becomes unstable, a small nucleus of oscillators is formed which synchronize at a frequency equal to $\bar\omega$ while the remaining part moves non-uniformly around the unit circle; finally, if $K$ reaches a critical value $K_c$, collective synchronization begins. Kuramoto proved that $K_c = 2/\pi g(0)$ in the limit  $N \rightarrow \infty$, where $g(\omega)$ is the distribution of natural frequencies, see~\cite{Strogatz_2000} for details.

The Kuramoto model refers to a fully (all-to-all) connected network. However, it is interesting to study collective synchronization also in the case of heterogeneous networks, because of this kind are those of particular importance in nature, social interactions and technological infrastructures. Considering the network topology, Eq.(~\ref{eq_1_kuramoto}) becomes:

\begin{equation}
\label{eq_4_kura_netw}
\dot\theta_i = \omega_i + \frac{K}{k_i} \sum_{j=1}^{N} A_{ij} \mbox{sin}(\theta_j - \theta_i) 
\end{equation}

where $A_{ij}$ is the adjacency matrix ($A_{ij} = 1$ if $\theta_i$ and $\theta_j$ are coupled, 0 otherwise) and $k_i$ is the number of connections of oscillator $\theta_i$. In this paper, only connected and symmetric networks ($A_{ij} = A_{ji}$) are considered.
The critical threshold value $K_c$ for the onset of synchronization has been theoretically calculated in the context of heterogeneous degree mean-field and the quenched mean-field theories.  The former predicts that $K_c$ is proportional to the ratio between the first two moments of the degree distribution, while the latter that it scales with the inverse of the largest  eigenvalue of the adjacency matrix~\cite{Rodrigues_2015}. In~\cite{Peron_2019} these predictions are compared for scale-free networks. Many other works also investigated the finite-size scaling of the order parameter~\cite{Moreno_2004,Lee_2005,Hong_2007,Hong_2013,Um_2014}.

%As far as we have been able to ascertain, there is no general theory that predicts for each type of network the value of the critical threshold $K_c$ for the onset of collective synchronization and, therefore, it is necessary to resort to numerical analysis. In some cases the system goes from a state of phase incoherence to one of coherence as soon as $K > K_c$ and it can be said that it undergoes a phase transition~\cite{English_2007}.

In this paper we focus on the study of the Kuramoto model for random networks. This type of networks are extensively studied both for their theoretical interest and for their statistical properties that make them a basic model in different  application fields. The novelty presented in this paper is the accurate numerical analysis of the synchronization process in the connectivity regime of random networks in the inhomogeneous Kuramoto model. In particular, the numerical results support the conjecture according to which the critical threshold of the coupling probability of oscillators above which they almost certainly synchronize is that of the network connectivity. Furthermore, to the best of our knowledge, this analysis has never been conducted for networks of the size of the largest network analyzed in this paper.

\section{Random networks}
The structure of a random network $G(N,p)$ of N nodes, also known as Erd\H{o}s-R\'{e}nyi,  is determined by the probability $0 \leq p \leq 1$ that a link exists between two randomly chosen nodes. The degree distribution of a random network, which is the probability that a randomly chosen node has degree $k$, is a binomial $P(k)=\binom{N-1}{k} p^k (1-p)^{N-1-k}$ and the average value of the degree is $\bar{k} = p (N-1)$. For $p < 1/N$ (subcritical region) the network is mainly composed of isolated nodes, tiny clusters and $\bar{k} < 1$. The largest cluster is a tree whose size is $N_G \sim \mbox{ln} N$. For $p = 1/N$ (critical point) $\bar{k}=1$, the network is composed of many small disconnected components, mainly trees, and $N_G \sim N^{2/3}$. For $1/N < p \leq \mbox{ln}N/N = p_c$ (supercritical region) $\bar{k} > 1$ and a giant component is formed which contains a finite part of the network and grows rapidly as $p$ increases. However, the network is not yet connected in this region. The probability $p_c$ is called connectivity threshold because  the network is supposed to became connected for $p$ just above $p_c$. For $p > p_c$ one has $\bar{k} \geq \mbox{ln}N$  and $N_G \sim N$. At the end of the connected region $p = 1$ and the network becomes fully (all-to-all) connected. See\cite{Barabasi_2016} for details.

\section{Method}
In this work, networks of size $N = 50, 100, 500$ and 1000 are considered. For each type of network analyzed 10 simulations are performed  and  for each of them the distribution of natural frequencies is a Gaussian with $\bar{\omega} = 0$ and $\sigma = 0.1$.
 Since isolated oscillators evolve independently, their dynamics do not affect the collective synchronization of those connected to the network.
Therefore, we consider the region with $p > p_c$. To study the limit $p \rightarrow p_c$, we set $p = p_c$ as parameter in the numerical library used to randomly generate the  networks~\cite{Networkx}, exclude disconnected configurations and accept only connected ones that  correspond to the case in which the giant component absorbs all the nodes of the network and has the minimum number of links.
The range of values of the coupling constant examined in this work is $0 \leq K \leq 1$ and we choose to divide it into 100 steps. For each value of $K$ the system of Eq.(~\ref{eq_4_kura_netw}) is numerically integrated up to 5000 time iterations of amplitude 0.1 s. The order parameter $R$ is calculated by averaging the last 1000 iterations in order to exclude values due to initial transient phase fluctuations. The value of $K_c$ was estimated by finite-size scaling analysis.

\section{Results}
\begin{figure*}
\begin{minipage}[h]{0.47\linewidth}
\begin{center}
\includegraphics[width=1\linewidth]{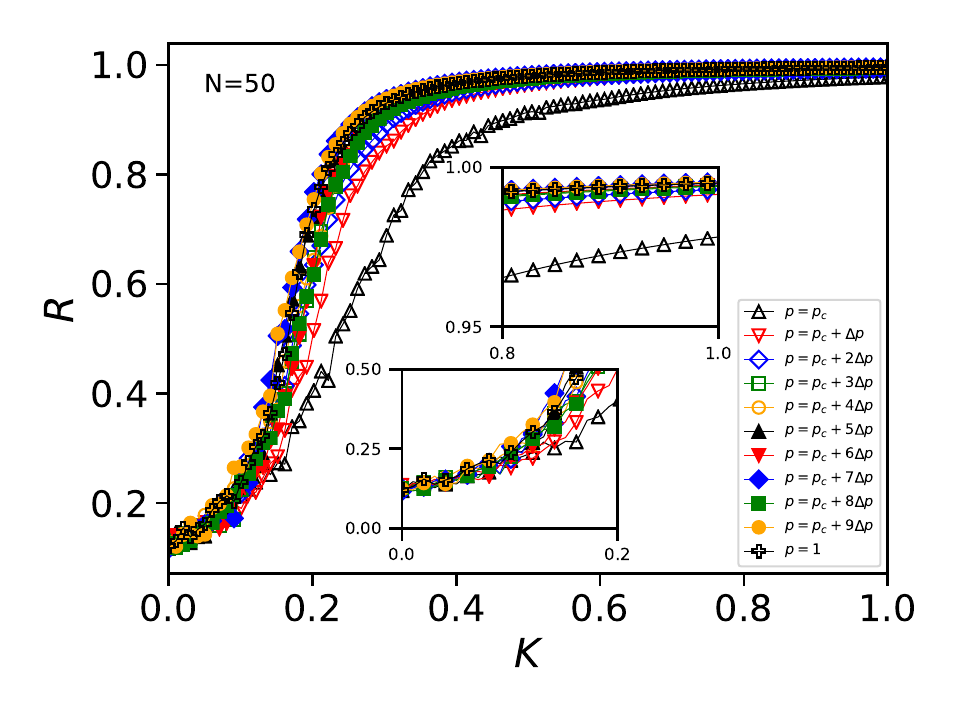}
%\caption{r1}
%\label{qwe1}
\end{center}
\end{minipage}
\hfill
%\vspace{0.2 cm}
\begin{minipage}[h]{0.47\linewidth}
\begin{center}
\includegraphics[width=1\linewidth]{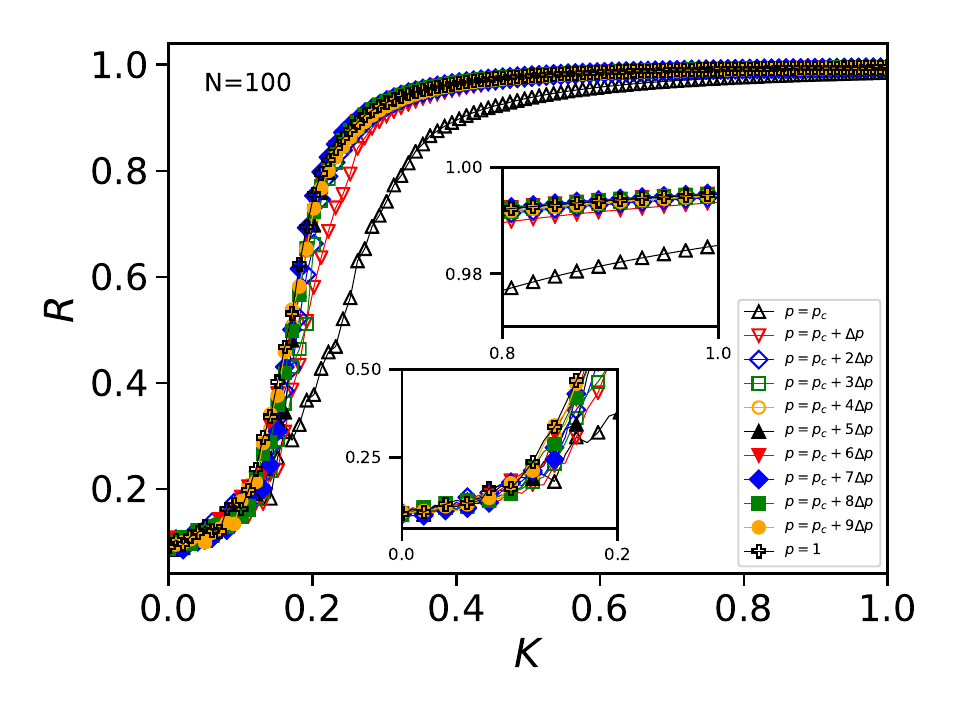}
%\caption{r1}
%\label{qwe1}
\end{center}
\end{minipage}
\vfill
%\vspace{0.2 cm}
\begin{minipage}[h]{0.47\linewidth}
\begin{center}
\includegraphics[width=1\linewidth]{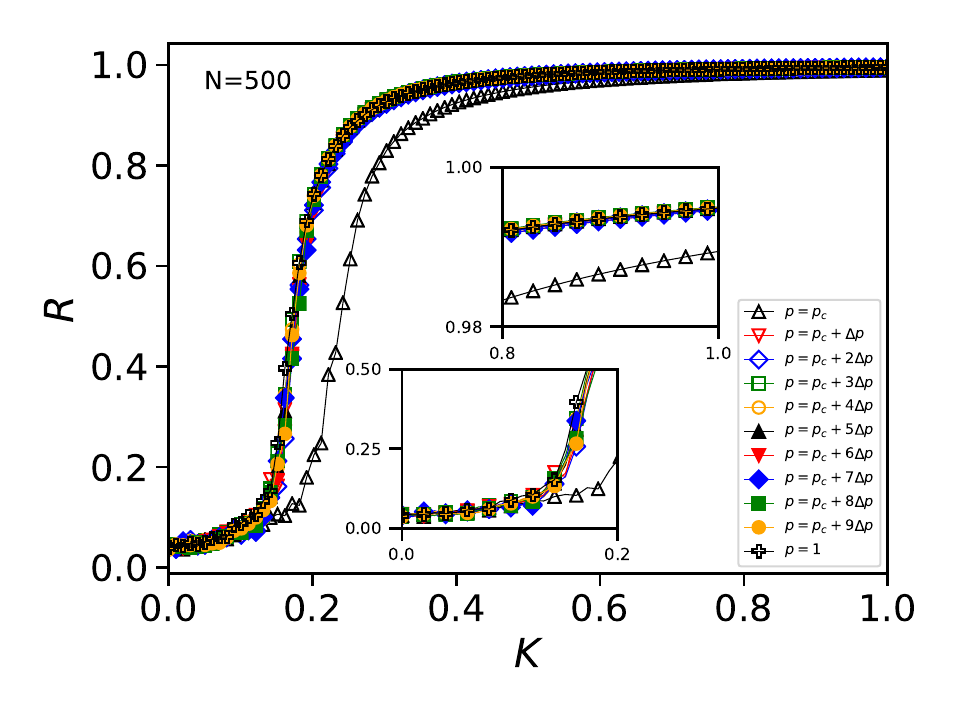}
%\caption{r1}
%\label{qwe1}
\end{center}
\end{minipage}
\hfill
\begin{minipage}[h]{0.47\linewidth}
\begin{center}
\includegraphics[width=1\linewidth]{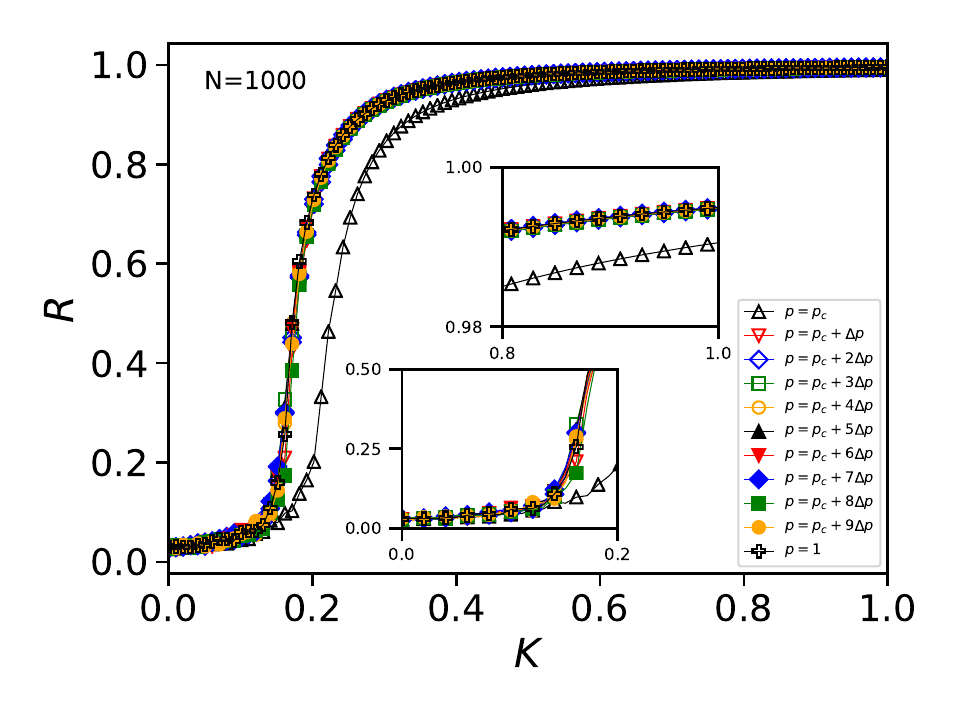}
%\caption{r1}
%\label{qwe1}
\end{center}
\end{minipage}
\caption{Order parameter $R$ as a function of the coupling constant $K$ for random networks of size $N = 50, 100, 500, 1000$. For each value of $K$ the system of Eq.(~\ref{eq_4_kura_netw}) is numerically integrated up to 5000 time iterations of amplitude 0.1 s. $R$ is calculated by averaging the last 1000 iterations. The figure shows $R(K)$ for different values of the probability $p$ that determines the structure of the network in the connected regime. For $p = p_c \simeq \mbox{ln}N/N$ the largest connected component absorbs all the nodes and has the minimum number of links. This is the configuration for which the network has the minimum level of connectivity. For $p = 1$ the maximum level of connectivity is achieved and every node is connected to all the others. The function $R(K)$ behaves roughly the same for $p > p_c$ while for $p = p_c$ it grows more slowly as $K$ increases. The figure shows in more detail the regions where the onset of synchronisation occurs and those where the oscillators are in phase coherence.  For $p > p_c$ the value of the critical threshold of the coupling constant for which synchronization occurs is $K_c \sim 0.16 \div 0.19$, while for $p = p_c$ it is $K_c \sim 0.21$.}
\label{fig3_r_vs_k_vs_p}
\end{figure*}

To study how the density of links affects synchronization we consider the probability increment $\Delta p = (1-p_c)/10$ and calculate the order parameter $R(K,p)$ for $p = p_c + n \Delta p$ for $n = 0,...,10$. The results are shown in Fig.~\ref{fig3_r_vs_k_vs_p}.

\begin{figure}
\includegraphics[width=1\linewidth]{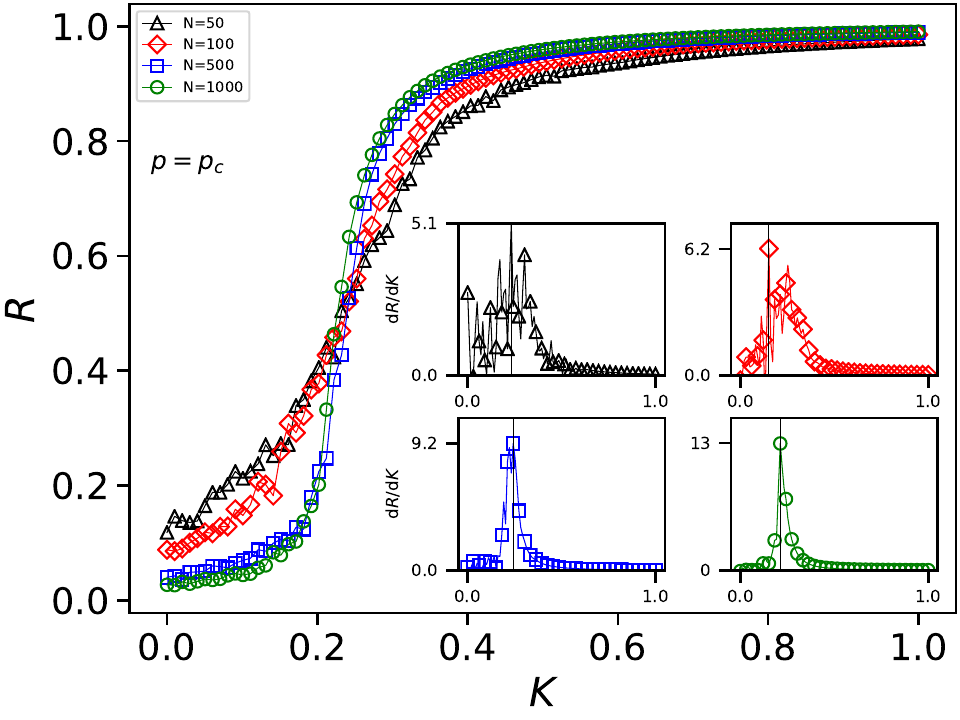}
\caption{The figure shows $R(K)$ when $p = p_c = \mbox{ln} N/N$ for all analyzed networks. As the size $N$ of the network increases the transition from $R \simeq 0$ to $R \simeq 1$ becomes abrupt and the oscillators have a phase transition. This is also evidenced by the inner figures which show that as $N$ increases  the derivative $dR/dK$ has a larger maximum and the variation around it becomes smaller.}
\label{fig4_r_vs_k_pc}
\end{figure}

What is observed is that the oscillators synchronize and reach the state of phase coherence for all values of $p$ examined in the connected region. The behavior of $R(K)$ is very similar for all values of $p > p_c$. For $p = p_c$, the system reaches phase coherence, although more slowly. As the size of the network grows, for $K > K_c$ the transition from state $R \simeq 0$ to $R \simeq 1$ is more and more sudden, as the shape of $R(K)$ shows, which means that the system has a phase transition. This also happens when $p = p_c$, as shown in more detail in Fig.~\ref{fig4_r_vs_k_pc}, where the derivative $dR/dK$ is also shown. It can be seen that as $N$ increases the maximum value of $dR/dK$ increases, while its variation around the maximum shrinks.

\begin{figure*}
\begin{minipage}[h]{0.47\linewidth}
\begin{center}
\includegraphics[width=1\linewidth]{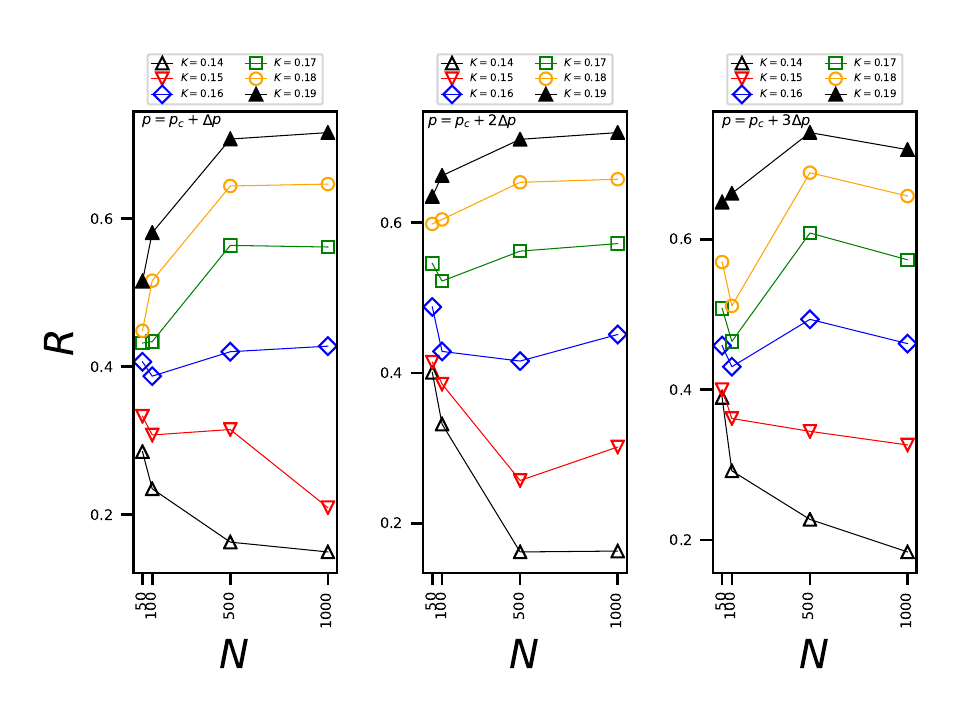}
%\caption{r1}
%\label{qwe1}
\end{center}
\end{minipage}
\hfill
%\vspace{0.2 cm}
\begin{minipage}[h]{0.47\linewidth}
\begin{center}
\includegraphics[width=1\linewidth]{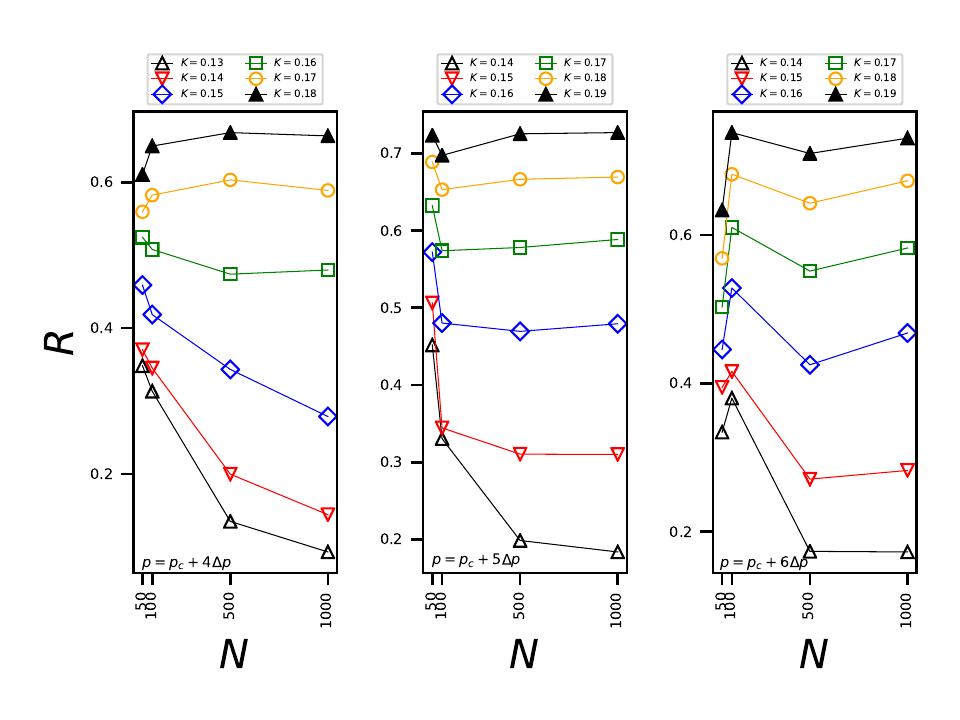}
%\caption{r1}
%\label{qwe1}
\end{center}
\end{minipage}
\vfill
%\vspace{0.2 cm}
\begin{minipage}[h]{0.47\linewidth}
\begin{center}
\includegraphics[width=1\linewidth]{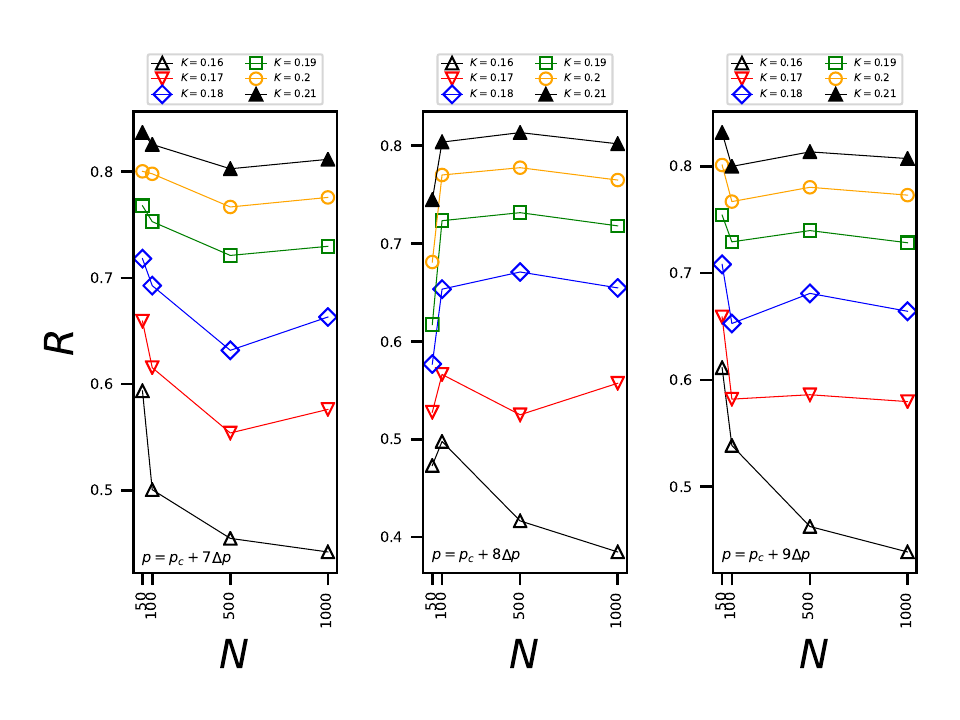}
%\caption{r1}
%\label{qwe1}
\end{center}
\end{minipage}
\hfill
\begin{minipage}[h]{0.47\linewidth}
\begin{center}
\includegraphics[width=1\linewidth]{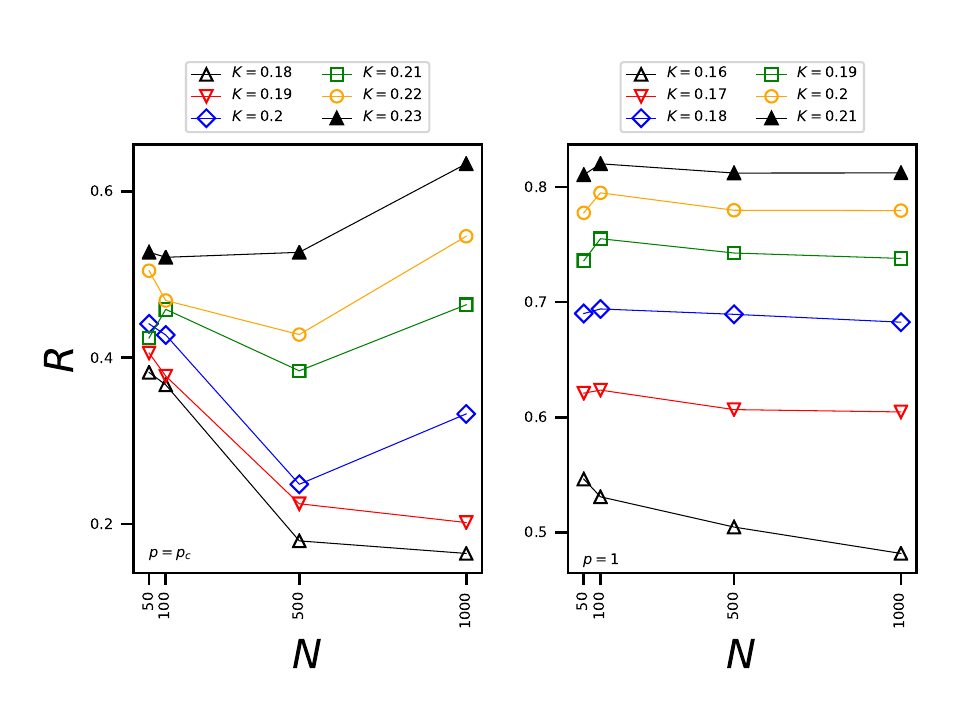}
%\caption{r1}
%\label{qwe1}
\end{center}
\end{minipage}
\caption{The figure shows $R(N,K)$ for all values of $p$. For $K < K_c$, R decays towards a residual of $O(1/\sqrt{N})$. For $K > K_c$, R saturates towards a value $R_{\infty} < 1$ with fluctuations of $O(1/\sqrt{N})$. For $K = K_c$ the variation of $R(N,K_c)$ is minimal. The best estimate is $K_c \sim 0.16 \div 0.19$ for $p > p_c$, and $K_c \sim 0.21$ for $p = p_c$.  }
\label{fig5_r_vs_N_vs_p}
\end{figure*}

To estimate $K_c$ we consider that for $K < K_c$ the oscillators do not synchronize and $R$ decays around a value of $O(1/\sqrt{N})$. For $ K> K_c$ $R$ increases and saturates at a value $R_{\infty} < 1$, although with fluctuations of $O(1/\sqrt{N})$. Therefore $K_c$ can be estimated numerically considering that at the onset of synchronization $R(N, K_c)$ has the smallest variation with $N$. Fig.~\ref{fig5_r_vs_N_vs_p} shows $R(N, K)$ for all values of $p$ examined in this work. For $p > p_c$ we have that $K_c \sim 0.16 \div 0.19$. This suggests that the onset of synchronization does not depend too much on the level of connectivity of the network itself. The theoretical value predicted by Kuramoto is $K_c \sim 0.16$. For $p = p_c$ the best estimate is $K_c \sim 0.21$, which confirms that for this network configuration the onset of synchronization is slower.

\section{Discussion}
The conjecture that the probability threshold above which synchronization is highly probable is $p_c = \mbox{ln}N/N$ has been considered by several authors. In~\cite{Ling_2019}, in the context of optimization of nonconvex cost functions, authors proved, for the homogeneous Kuramoto model, that if $p \gg \mbox{ln}N/N^{1/3}$ the oscillators synchronize with high probability for $N \rightarrow \infty$. In~\cite{Kassabov_2022}, with arguments from spectral graph theory, authors proved that the same occurs if $p \gg \mbox{ln}^2 N/N$. The issue has also been addressed in the context of graphon theory~\cite{Lovasz_2012,Borgs_2019,Medvedev_2014_a,Medvedev_2014_b,Medvedev_2019}. In~\cite{Nagpal_2024} authors proved, still for the case of the homogeneous model and for $N \rightarrow \infty$,  that the oscillators reach phase synchronization with high probability if $p > p_c$. They performed numerical simulations for networks of size $N \leq 100$.
In this work the above conjecture has been proved numerically even in the more general case of inhomogeneous Kuramoto model and for larger network sizes.

\section{Conclusion}
We have numerically studied the collective synchronization of coupled phase oscillators in the Kuramoto model in the connected regime of random networks of sizes $N = 50, 100, 500$ and 1000. The oscillators synchronize and reach phase coherence for all connectivity levels of the networks. The behavior of the order parameter $R$ as a function of the coupling constant $K$ does not show significant discrepancies between different levels of link density. For the limiting case in which oscillators are coupled with probability around the connectivity threshold of the network, synchronization still occurs, although in a more attenuated way. As the size of the network increases, the transition from the incoherent to the coherent phase state occurs more and more abruptly and the oscillators have a phase transition. The critical threshold value  for the onset of synchronization varies in the range $K_c \sim 0.16 \div 0.19$  for all the connectivity levels of the networks. In the limiting case when the network goes from disconnection to connection  $K_c \sim 0.21$.
A random network of a certain size can be obtained by perturbing a complete network of the same size by randomly eliminating links. Therefore, from this analysis it can be deduced that a random perturbation of a complete network of oscillators coupled according to the Kuramoto model does not affect the synchronization as long as the perturbation does not disconnect the network.

\ack
The computing resources and the related technical support used for this work have been provided by CRESCO/ENEAGRID High Performance Computing infrastructure and its staff~\cite{Iannone_2019}. CRESCO/ENEAGRID High Performance Computing infrastructure is funded by ENEA, the Italian National Agency for New Technologies, Energy and Sustainable Economic Development and by Italian and European research programmes, see http://www.cresco.enea.it/english for information


\section{References}

\begin{thebibliography}{30}

\bibitem{Banfalvi_2011} 
Banfalvi G 2011 {\it Methods Mol Biol.} \textbf{761} 1

\bibitem{Manukyan_2011} 
Manukyan A, Abraham L, Dungrawala H and Schneider B L 2011 {\it Methods Mol Biol.} \textbf{761} 173

\bibitem{Ziping_1993}
Ziping J and McCall M 1993 {\it J. Opt. Soc. Am.} \textbf{10} 155

\bibitem{Kourtchatov_1995}
Kourtchatov S Y, Likhanskii V V, Napartovich A P, Arecchi F T and Lapucci A 1995 {\it Phys. Rev.} A \textbf{52} 4089

\bibitem{Takemura_2021}
Takemura N, Takata K, Takiguchi M and Notomi M 2021 {\it Sci. Rep.} \textbf{11} 8587

\bibitem{Wiesenfeld_1998}
Wiesenfeld K, Colet P and Strogatz S H 1998 {\it Phys. Rev.} E \textbf{57} 15631

\bibitem{Shahal_2020}
Shahal S, Wurzberg A, Sibony I, Duadi H, Shniderman E, Weymouth D, Davidson N and Fridman M 2020 {\it Nat. Commun.} \textbf{11} 3854

\bibitem{Vandermeer_2021}
Vandermeer J, Hajian-Forooshani Z, Medina N and Perfecto I 2021 {\it R. Soc. Open Sci.} \textbf{8} 210122

\bibitem{Winfree_1967}
Winfree A T 1967 {\it J. Theoret. Biol.} \textbf{16} 15

\bibitem{Kuramoto_1984}
Kuramoto Y 1984 {\it Chemical Oscillations, Waves and Turbulence} vol~19 (Berlin: Springer) 

\bibitem{Strogatz_2000}
Strogatz S H 2000 {\it Physica} D{\it: Nonlinear Phenomena} \textbf{143} 1

\bibitem{Rodrigues_2015}
Rodrigues F A, Peron T, Ji P and Kurths J 2015 {\it Physics Reports} \textbf{610} 1

\bibitem{Peron_2019}
Peron T, Messias B, Mata A S, Rodrigues F A and Moreno Y 2019 {\it Phys. Rev} E \textbf{100} 042302

\bibitem{Moreno_2004}  
Moreno Y and Pacheco A F 2004 {\it Europhysics Letters} \textbf{68} 603

\bibitem{Lee_2005}
Lee D-S 2005 {\it Phys. Rev} E \textbf{72} 026208

\bibitem{Hong_2007}
Hong H, Park H and Tang L-H 2007 {\it Phys. Rev} E \textbf{76} 066104

\bibitem{Hong_2013}
Hong H, Um J and Park H 2013 {\it Phys. Rev} E \textbf{87}  042105

\bibitem{Um_2014}
Um J, Hong H and Park H 2014 {\it Phys. Rev} E \textbf{89} 012810

\bibitem{English_2007}
English L Q 2007 {\it European Journal of Physics} \textbf{29} 143

\bibitem{Barabasi_2016}
 Barab\'{a}si A 2016 {\it Network science} (Cambridge University Press)

\bibitem{Networkx}
Hagberg A A, Schult D A and Swart P J 2008 {\it Proc. of the 7th Python in Science Conference (SciPy2008)} (Pasadena, California) p~11 doi: 10.25080/PFVC8793

\bibitem{Ling_2019}
Ling S, Xu R and Bandeira A S 2019 {\it SIAM J. Optim.} \textbf{29} 1879

\bibitem{Kassabov_2022}
Kassabov M, Strogatz S H and Townsend A 2022 {\it Chaos: An Interdisciplinary Journal of Nonlinear Science} \textbf{32} 093119

\bibitem{Lovasz_2012}
Lov\'{a}sz L 2012 {\it American Mathematical Soc.} \textbf{60} 

\bibitem{Borgs_2019}
Borgs C, Chayes J, Cohn H and Zhao Y 2019 {\it Transactions of the American Mathematical Society} \textbf{372} 3019

\bibitem{Medvedev_2014_a}
Medvedev G S 2014 {\it SIAM Journal on Mathematical Analysis} \textbf{46} 2743

\bibitem{Medvedev_2014_b}
Medvedev G S 2014 {\it Archive for Rational Mechanics and Analysis} \textbf{212} 781

\bibitem{Medvedev_2019}
Medvedev G S 2019 {\it Communications in Mathematical Sciences} \textbf{17} 883

\bibitem{Nagpal_2024}
Nagpal S V, Nair G G, Strogatz S H and Parise F 2024 Synchronization in random networks of identical phase oscillators: A graphon approach arXiv:2403.13998 [math.DS]

\bibitem{Iannone_2019}
Iannone F \etal 2019 {\it Proc. Int. Conf. on High Performance Computing  Simulation (HPCS)} (Dublin, Ireland. Publisher: IEEE) p~1051--1052 doi: 10.1109/HPCS48598.2019.9188135

\end{thebibliography}
\end{document}